\documentclass[twoside]{report}
\usepackage{svcon2e}
\usepackage{amsmath,amsthm,amsfonts,latexsym,amssymb}

\outer\def\proof{\smallbreak\noindent{\bf Proof.}\enspace}

\def\endproof{$\square$}

\def\re{{\Re e}\,}

\def\C{\mathbb {C}}
\def\R{\mathbb{R}}

\def\dom{{\rm dom\,}}

\def\ch{{\cal H}}

  \newcommand{\mynewtheorem}[3]{\newenvironment{#1}{\subsection{#2} #3}{}}
\mynewtheorem{claim}{Claim}{\it}
\mynewtheorem{plem}{Purity Lemma}{\it}

\newtheorem{thm}{Theorem}[section]
  \newtheorem{lem}[thm]{Lemma}

\begin{document}

\makeatletter
\renewcommand{\@oddhead}{\thepage\hskip1.5cm{\sl
Interaction representation method
in quantum optics
}\hfil}
\makeatother

\chapter{Interaction representation method for Markov master equations
in quantum optics
}
\chapterauthors{Alexander M.~Chebotarev,
Julio C.~Garcia and
Roberto B.~Quezada
}
\footnotetext{
Partially supported by INTAS grant 99--00545 and by
CONACYT grant 28520E; will be published in:
{\it Trends in Mathematics, Stochastic Analysis and
Mathematical Physics}, ANESTOC, Proc. of the Fourth International
Workshop, Birkhauser, (R.Rebolledo ed.), Boston, 2001.
}
\abstract{
Conditions sufficient for
a quantum  dynamical semigroup
(QDS) to be unital are proved for a
 class of problems in quantum optics with Hamiltonians which
are self-adjoint polynomials
of any finite order in creation and
annihilation operators.
The order of  the Hamiltonian
may be higher than the order of completely positive part
of  the  formal generator of a QDS.

The unital property of a minimal quantum  dynamical semigroup implies the
uniqueness of the solution of the corresponding
Markov master equation in the
class of quantum dynamical semigroups and, in the corresponding
representation, it ensures preservation
of the trace or unit operator.
We recall that only in the unital
 case
the formal generator of MME determines uniquely the
corresponding QDS.
}

\section*{Introduction}

Numerical experiments with the remarkable quantum trajectories algorithm
 \cite{SB96} (Schack \& Brun `97),
which solves the Markov  master equation (MME)
by the Monte-Carlo method, show
that the numerical code forces
a solution to be unital even in cases, when the exact minimal solution
does not preserves the trace of initial state or
unit initial operator, and the corresponding Poisson
process explodes on finite time intervals.
Unfortunately, the mathematical theory of MME still does not
work for many basic equations
in quantum physics. In this paper we extend the approach
developed in \cite{CF98} to  MMEs in quantum optics.

The formal generator ${\cal  L}(\cdot)$ of a quantum
dynamical semigroup  on von Neumann
algebra ${\cal  B}({\cal  H})$ of all bounded
operators in a separable Hilbert space ${\cal  H}$ is called
{\sl regular} if it defines the semigroup in an unambiguous way.
Following \cite{L76}, we assume that the
coefficients of the generator ${\cal  L}(\cdot)$
are a densely defined symmetric Hamiltonian operator $H$
and a completely positive map
$\Phi(\cdot)$.
The structure of the generator ${\cal  L}(\cdot)$ is similar to the structure
of the classical Kolmogorov--Feller equation, and similarly to the
classical case, under rather general assumptions, MME has the minimal
solution called  a {\sl minimal quantum dynamical semigroup} (QDS).
Moreover, if the minimal QDS is
unital, it is the unique solution of the corresponding MME; this
implies regularity of the formal generator ${\cal  L}(\cdot)$.

In the present paper we suggest a new test for regularity of MME
with the formal generator ${\cal  L}(\cdot)$.
The idea consists in a suitable choice
of some $\Lambda$-pair  consisting of
a "reference" operator $\Lambda$ and
"interaction" part $H_{int}=\Lambda+H_{s.a.}$,
which generates the interaction representation for
the Hamiltonian $H=H_s+H_{s.a.}=(H_s-\Lambda)+(\Lambda+H_{s.a.})$,
where $H_{s}$ and $H_{s.a.}$ are symmetric and self-adjoint
components of $H$.

We call the pair $(\Lambda, \,H_{s.a.})$
a $\Lambda$-{\sl pair} for the generator  ${\cal  L}(\cdot)$,
if the positive self-adjoint $\Lambda$ is a {\sl reference operator} for
the reduced generator ${\cal  L}_0(\cdot)$, i.e.
\begin{gather*}
{\cal  L}_0(X)=\Phi(X)-G_0^*X-XG_0,\quad G_0=\frac{1}{2}\Phi(I)
+i(H_s-\Lambda),\\
{\cal  L}_0(\Lambda)\le c\Lambda, \quad I\le\Phi(I)\le\Lambda,
\end{gather*}
and $H_{s.a.}$ and $H_s$ are such that
\begin{gather*}
-\mu H_{int}^{\varepsilon }\le H_{s.a}\le \mu H_{int}^{\varepsilon },\quad
0\le H_{int}\le \nu \Lambda,
\\
\Phi(H_{s.a.})\le c_1 \Lambda,\quad
\Phi(H_{int}^{\varepsilon })\le c_2 \Lambda,
\end{gather*}
for some $c,\,\mu,\,\nu\ge0$ and $\varepsilon\in(0,1)$.
Under these algebraic assumptions, together with some domain and continuity
conditions
which will be specified in Sections 1 and 2,
we prove that the minimal dynamical semigroup with the formal
generator ${\cal  L}(\cdot)$ is unital.

The paper is organized as follows.
In Section 1, we discuss the construction of a minimal
solution for the Markov master equation
with time-dependent coefficients and prove new rather general
conditions sufficient for the conservativity of a minimal solution.
In Section 2, we discuss the properties of the interaction
representation of MME
and introduce definitions of a reference family and $\Lambda$-pair.
In Section 3, we describe the properties
of generators of MME in quantum optics, and present
basic examples of violation of the unital property.
In Section 4
several physical examples of nontrivial MME are considered which
in the most cases are regular.

The main conclusion of the paper
is rather simple: for the class of CP-maps  $\Phi(\cdot)$
in quantum optics, the minimal QDS is unital if the
Hamiltonian $H$ is  a s.a. operator.

\section{Conditions sufficient for conservativity}

Consider the Markov master equation with time-dependent
generator
\begin{gather}
\frac{\partial }{\partial t} P_{\tau,t}(B)={\cal  L}_t(P_{\tau,t}(B)),\quad
P_{\tau,t}(B)|_{\tau=t}=B,
\\
{\cal  L}_t(B)=\Phi_t(B)-G^*_tX -XG_t,\quad G_t=\frac{1}{2}\Phi_t(I)+iH_t.
\notag
\end{gather}
and assume that there exists strongly continuous
contractive evolution system
$W_{s,t}\in{\cal  B}({\cal  H})$  $(t\ge s)$ such that
$$
W_{s,\tau}W_{\tau,t}=W_{s,t},\quad
\frac{\partial }{\partial t}W_{s,t}=-W_{s,t}G_t,\quad
\frac{\partial }{\partial s}W_{s,t}=G_sW_{s,t}.
$$
For simplicity, we assume that all the generators $G_t$ have a joint core
${\cal  D}_N\subseteq \dom G_t^N$ for some $N\ge 1$  such that
 the following preliminary {\sl domain assumptions} are
fulfilled
\begin{equation}
W_{s,t}:{\cal  D}_N\to {\cal  D}_N \subseteq
\dom \Phi_{*,t}(I)[\,\cdot\,],\quad
G_s^*+G_s=\Phi_s(I)  \; {\rm on}\;{\cal  D}_N,
\end{equation}
and that there exists some positive  self-adjoint operator
$\Lambda: \Lambda \ge \Phi_t(I)\ge I$ such that
the preliminary {\sl continuity conditions} are fulfilled:
\medskip

\noindent
{\it the family of CP-maps $\Lambda^{-1/2}\Phi_t(\cdot)\Lambda^{-1/2}$
is bounded, normal and ultraweakly continuous in $t$, and on the other
hand, for any $\psi\in {\cal  D}_N$, the family of vectors
$\psi(s,t)=\Lambda^{1/2}W_{s,t}\psi$ belongs to $L_2^{loc}(\R_+,{\cal  H})$
in variable  $s$
and is norm-continuous in  variable $t$}.

\medskip

Under these domain and continuity conditions,
for any bounded strongly continuous family of operators
$X_s$, the family of quadratic forms
\begin{equation}
\Phi_s(X_s)[W_{s,t}\psi]
\quad \forall \psi\in{\cal  D}_N
\end{equation}
belongs to $L_1^{loc}(\R_+)$  in variable $s$ and is continuous in $t$.

Hence the sequence of CP-maps
\begin{align*}
P^{(0)}_{\tau,t}(B)&=W^*_{\tau,t}BW_{\tau,t}\stackrel{def}{=}V_{\tau,t}(B),\\
P^{(n+1)}_{\tau,t}(B)&\stackrel{def}{=}V_{\tau,t}(B)+
\int_\tau^t\,ds \,V_{s,t}\Phi_sV_{\tau,s}P^{(n)}_{\tau,s}(B)
\end{align*}
is well-defined as a sequence of bounded operators corresponding
to the sequence of densely defined and uniformly bounded quadratic forms.
Indeed, $V_{\tau,t}(I)\le I$, and the identity
\begin{equation}
V_{\tau,t}(I)=I-\int_\tau^t\,ds \,V_{s,t}\Phi_s(I)=
I-\int_\tau^t\,ds \,V_{\tau,s}\Phi_s(I)
\end{equation}
readily shows that the sequence $P^{(n)}_{\tau,t}(B)$ is
uniformly bounded:
$$
||P^{(n)}_{\tau,t}(B)||\le ||B||.
$$
Moreover, it increases monotonically if $B\in {\cal  B}_+({\cal  H})$,
and defines the least upper bound:
$$
P^{min}_{s,t}(B)={\rm l.u.b.\,}P^{(n)}_{s,t}(B).
$$
This construction is analogous to the construction of
the minimal solution for the Markov master equation with constant
coefficients \cite{Ch91}--\cite{Ch92}; it was discussed in details
in \cite{CGQ98}.

The identity (1.4) implies that
\begin{align*}
&P^{(1)}_{\tau,t}(I)=V_{\tau,t}(I)+\int_\tau^t\,ds\,
V_{s,t}\Phi_sV_{\tau,s}(I)\\
&=V_{\tau,t}(I)+\int_\tau^t\,ds\,
V_{s,t}\Phi_s(I)-\int_\tau^t\,ds_1\,
V_{s_1,t}\Phi_{s_1}\int_\tau^{s_1}\,ds_2\,V_{s_2,s_1}\Phi_{s_2}(I)
\\
&=I-\int_\tau^t\,ds_1\,
V_{s_1,t}\Phi_{s_1}\int_\tau^{s_1}\,ds_2\,V_{s_2,s_1}\Phi_{s_2}(I).
\end{align*}
Similarly, by using sequentially the identity (1.4), we obtain
\begin{gather}
P^{(n)}_{\tau,t}(I)=
I-\Delta^{(n+1)}(\tau,t),\notag\\
\Delta^{(n)}(\tau,t)\stackrel{def}{=}\int_\tau^t\,ds_1\,
V_{s_1,t}\Phi_{s_1}\dots \int_\tau^{s_{n-1}}\,ds_{n}\,
V_{s_{n},s_{n-1}}\Phi_{s_{n}}(I).
\end{gather}
Hence, $P^{(n)}_{\tau,t}(I)\to I$ strongly as $n\to\infty$,
if and only if $\Delta_n(s,t)\to0$ weakly.

To prove a condition sufficient for the
minimal solution of the Markov master equation (1.1)
to be unital, let us consider
an estimate for the integral of the
operator $P^{(n)}_{\tau,t}\Phi_{\tau}(I)$:
\begin{gather*}
\int_\tau^t\,ds\,P^{(n)}_{s,t}\Phi_{s}(I)
=\int_\tau^t\,ds\,V_{s,t}\Phi_s(I)+\dots\\
+\int_\tau^t\,ds\,\int_s^t\,ds_1\,V_{s_1,t}\Phi_{s_1}\dots
\int_s^{s_{n-1}}\,ds_n\,V_{s_n,s_{n-1}}\Phi_{s_n}V_{s,s_{n}}\Phi_{s}(I).
\end{gather*}
In the last multiple integral, the variables $s_k$ take
greater values  then $s$, i.e.
$\tau\le s\le s_n\le\dots\le s_1$. Hence by changing the order of
integration,  we have
\begin{gather}
\int_\tau^t\,ds\,P^{(n)}_{s,t}\Phi_{s}(I)
=\int_\tau^t\,ds\,V_{s,t}\Phi_s(I)+\dots
\notag\\
+\int_\tau^t\,ds_1\,V_{s_1,t}\Phi_{s_1}\dots
\int_s^{s_{n-1}}\,ds_n\,V_{s_n,s_{n-1}}\Phi_{s_n}
\int_\tau^{s_n}\,ds\, V_{s,s_{n}}\Phi_{s}(I)
\end{gather}
The last integral in (1.6) can be rewritten as the last integral
in (1.5) in notation $s\to s_{n+1}$.
By comparing Eqs.~(1.5) and (1.6)
and passing to the least upper bound in $n$,
we obtain the following important
equality:
\begin{equation}
\int_\tau^t\,ds\,P^{min}_{s,t}\Phi_{s}(I)=
\sum_{n=1}^\infty
\Delta^{(n)}(\tau,t),
\end{equation}
where  the monotone sequence of bounded positive operators
$\Delta^{(n)}(\tau,t)$ (see (1.4)) converges to $0$ if and only if
the integral in the left-hand side is a densely defined
operator.
To make  rigorous the above algebraic considerations of integrals,
we must impose additional assumptions on domains and continuity.

In the sequel we assume that the CP-map $\Phi_t(\cdot)$ is such that
for each $t\in\R_+$ the map
$A_t(\cdot)=\Lambda^{-1/2}\Phi_t(\cdot)\Lambda^{-1/2}$ is bounded and normal.
In fact,
the boundedness follows from the inequality
$\Lambda\ge \Phi_t(I)$. The Kraus theorem  \cite{Kr}
implies that any normal
bounded CP-map $A_t(X)$ on ${\cal  B}({\cal  H})$ (${\cal  H}$ is a separable
Hilbert space) can be represented
as the sum $A(X)=\sum_k A_k^*(t)XA_k(t)$,
$\sum_k A_k^*(t)A_k(t)\in{\cal  B}({\cal  H})$.
This ensures a canonical representation of
unbounded CP-map $\Phi_t(\cdot)$ \cite{Ga99}:
$$
\Phi_t(X)=\sum_k \Phi_k^*(t)X\Phi_k(t),\quad
\Phi_k(t)=A_k(t)\Lambda_t^{1/2},
$$
where $\sum_k A_k^*(t)A_k(t)\in{\cal  B}({\cal  H})$.
To study  conditions  sufficient for the minimal solution to be unital,
we must extend the domain and continuity assumptions.
We assume that for some $N\ge 2$ the operators $\Lambda^{1/2}\Phi_k(t)$
are densely defined,
\begin{equation}
{\cal  D}_N\subseteq \dom \Lambda^{1/2}\Phi_k(t),
\end{equation}
and $\Lambda^{1/2}\Phi_k(s)W_{s,t}\psi\in L_2^{loc}(\R_+,{\cal  H})$
in variable $s$ and norm-continuous in $t$.
Thus the inequality (1.7) justifies the following assertion.

\begin{thm}
Under the domain and continuity
assumptions, if the
domain of the operator $\int_\tau^t\,ds\,P^{min}_{s,t}\Phi_{s}(I)
$ is dense in ${\cal  H}$, then the minimal solution
of the Markov master equation {\rm (1.1)} is unital.
\end{thm}

Since the sequence $\Delta^{(n)}(\tau,t)$ is positive and
decreases monotonically, the sum
$$
C=\sum_{n=1}^\infty a_n\Delta^{(n)}(\tau,t), \quad a_n\ge 0,\quad
\sum_n a_n=\infty
$$
converges to a densely defined operator only if $\Delta^{(n)}(\tau,t)$
converges to $0$. The series which correspond to this sum
with $a_n=n^{-1}$ can be
represented as an integral of the minimal solutions of MMEs
with the generators regularized as in \cite{D79}:
$$
{\cal  L}_{t,\lambda}(B)
=\lambda\Phi_t(B)-G^*_tX -XG_t,\quad \lambda\in (0,1].
$$
More precise, the series, representing the minimal solution
of the equation
$$
\frac{\partial }{\partial t} P_{s,t}^{(\lambda)}(B)=
{\cal  L}_{t,\lambda}(P_{s,t}^{(\lambda)}(B)),\quad
P_{s,t}^{(\lambda)}(B)|_{s=t}=B
$$
is the following:
$$
P^{(\lambda)}_{\tau,t}(B)=V_{\tau,t}(B)+\sum_{n=1}^\infty
\lambda^n\int_\tau^t\,ds_1\,
V_{s_1,t}\Phi_{s_1}
\dots \int_\tau^{s_{n-1}}\,ds_{n}\,\Phi_{s_n}V_{\tau,s_n}(B).
$$
This identity and definition (1.5) imply
\begin{equation}
\int_0^1\,d\lambda\,\int_\tau^t\,ds \,
P^{(\lambda)}_{s,t}\Phi_s(I)=\sum_1^\infty\frac{1}{n}\Delta^{(n)}(\tau,t).
\end{equation}
Therefore, the following assertion is true.

\begin{thm}
Assume that the domain and continuity
conditions are fulfilled. If the operator
$$
\widehat C=\int_0^1\,d\lambda\,\int_\tau^t\,ds\,P^{(\lambda)}_{s,t}\Phi_s(I)
$$
is densely defined in ${\cal  H}$, the minimal solution of the Markov
master equation {\rm (1.1)} is unital.
\end{thm}

Let us consider a priori bounds for the operator
$P_{s,t}\Phi_s(I)$.

\section{A priori bounds}
\setcounter{equation}{0}

Assume that there exists a smooth family of positive
self-adjoint operators $\Lambda_t\ge \Phi_t(I)\ge I$ and
$N\ge 2$ such that for any $\psi\in{\cal D}_N\subseteq\dom\dot\Lambda_t$
and
\begin{equation}
{\Phi}_{*,t}(\Lambda_t)[\psi]-2\Re e\,(G_t\psi,\Lambda_t\psi)
-(\psi,\dot\Lambda_t\psi)\le c_t||\Lambda_t^{1/2}\psi||,
\end{equation}
(cf. \cite{CF98} and \cite{Ho96}),
where $c_t\in L_1^{loc}(\R_+)$, $c_t\ge0$.
Such operator family is called a family of {\sl reference}
operators.
We assume that the family of operators $\Lambda^{-1/2}\Lambda_t
\Lambda^{-1/2}$, with the previously defined operator $\Lambda$,
is densely defined on ${\cal  H}$ and admits a continuation on the
whole space ${\cal  H}$  which is uniformly bounded
and strongly continuous.
Let us prove that condition (2.1) ensures the a priori estimate
\begin{equation}
P^{min}_{\tau,t}(\Lambda_\tau)\le\Lambda_t e^{\int_\tau^t\,ds\,c_s}.
\end{equation}
This estimate can easily be proved by induction. Indeed, for all
$0\le\tau\le t$ we have
\begin{align*}
\frac{\partial }{\partial \tau}V_{\tau,t}\Lambda_\tau&=
\frac{\partial }{\partial \tau}W^*_{\tau,t}\Lambda_\tau W_{\tau,t}
= W^*_{\tau,t}(G_\tau^*\Lambda_\tau+\Lambda_\tau G_\tau+
\dot \Lambda_\tau) W_{\tau,t}
\\
&\ge W^*_{\tau,t}(\Phi_\tau(\Lambda_\tau)+
c_\tau \Lambda_\tau) W_{\tau,t}\ge  c_\tau
W^*_{\tau,t}\Lambda_\tau W_{\tau,t}=c_\tau V_{\tau,t}\Lambda_\tau,
\end{align*}
Hence, by solving this terminal differential inequality, we obtain
$$
P^{(0)}_{\tau,t}(\Lambda_\tau)=
V_{\tau,t}\Lambda_\tau \le\Lambda_te^{\int_\tau^t\,ds\,c_s}.
$$
Assume that $P^{(n)}_{\tau,t}(\Lambda_\tau)
\le\Lambda_te^{\int_\tau^t\,ds\,c_s}$ and let us prove this inequality
for  $P^{(n+1)}_{\tau,t}(\Lambda_\tau)$.

From the recurrent definition of  $P^{(n+1)}_{\tau,t}(\Lambda_\tau)$
and assumption (2.1) we have
\begin{align*}
P^{(n+1)}_{\tau,t}(\Lambda_\tau)&=
W^*_{\tau,t}\Lambda_\tau W_{\tau,t}+
\int_\tau^t\,ds\,V_{s,t}\Phi_sP^{(n)}_{\tau,s}(\Lambda_\tau)
\\
&\le W^*_{\tau,t}\Lambda_\tau W_{\tau,t}+
\int_\tau^t\,ds\,e^{\int_\tau^s\,dr\,c_r}V_{s,t}\Phi_s(\Lambda_s)
\\
&\le W^*_{\tau,t}\Lambda_\tau W_{\tau,t}+
\int_\tau^t\,ds\,e^{\int_\tau^s\,dr\,c_r}V_{s,t}(c_s\Lambda_s+
G^*_s\Lambda_s+\Lambda_sG_s+\dot\Lambda_s)
\\
&=V_{\tau,t}\Lambda_\tau +
\int_\tau^t\,ds\,\frac{\partial }{\partial s}\biggl(e^{\int_\tau^s\,dr\,c_r}
V_{s,t}\Lambda_s\biggr)=\Lambda_te^{\int_\tau^t\,ds\,c_s}.
\end{align*}
This estimate readily implies that the operator
$P_{\tau,t}^{min}(\Lambda_\tau)$ is densely defined,
$\dom \Lambda\subseteq \dom P_{\tau,t}^{min}(\Lambda_\tau)$,
 and hence the unital property holds.
Therefore, the following assertion holds true.

\begin{thm}
Let the domain and continuity assumptions be fulfilled and there exist
a reference family $\Lambda_t$. Then
the minimal solution
of Eq.~{\rm (1.1)} is unital.
\end{thm}

 Assume that for a formal generator ${\cal  L}(\cdot)$
 with constant coefficients
there exists some constant reference operator $\Lambda$, i.e.
$$
{\cal  L}(\Lambda)\le c\Lambda, \quad {\cal  L}(B) =\Phi(B)-G^*B-BG,
$$
where $G=\frac{1}{2}\Phi(I)+iH$, $H=H_s+H_{s.a.}$, and let
$H_{int}=H_{s.a.}+\Lambda$ be a self-adjoint operator.
Then inequality (2.1) holds for ${\cal  L}_t(B) =\Phi_t(B)-G_t^*B-BG_t$
and the reference family
$\Lambda_t=U_t^*\Lambda U_t$, where
$U_t=e^{itH_{int}}$
$$
G_t=\frac{1}{2}U_t^*\biggl(\frac{1}{2}\Phi(I)+i(H_s-\Lambda)\biggr)U_t,\quad
\Phi_t(B)= U_t^*\Phi(U_tBU_t^*) U_t.
$$

Let us discuss an opportunity to use some fixed reference
operator $\Lambda$ for problems with time-dependent coefficients,
which arise in the interaction representation.

Let $\Lambda\ge0$ and $H_{s.a.}$ be self-adjoint operators such that
the sum $H_{int}=H_{s.a.}+\Lambda$ is positive and self-adjoint,
and there exist $\mu,\,\nu\ge0$ and $\varepsilon \in(0,1)$ such that
\begin{equation}
-\mu H_{int}^{\varepsilon }\le H_{s.a}\le \mu H_{int}^{\varepsilon },\quad
0\le H_{int}\le \nu \Lambda.
\end{equation}
Note that for any positive self-adjoint operator $X$ and
$\varepsilon\in(0,1] $, we have
$X^{\varepsilon }\le I+X$, since $\lambda^\varepsilon\le
1+\lambda$ for any positive $\lambda$, and hence
$$
X^{\varepsilon }=\int\,\lambda^\varepsilon\,E_X(d\lambda)\le
\int\,(1+\lambda)\,E_X(d\lambda),
$$
where $E_X(d\lambda)$ is the spectral family of the operator $X$.
Then
\begin{align*}
\Lambda_t&=U_t\Lambda U^*_t=U_t(H_{int}-H_{s.a.})U^*_t
\\
&=H_{int}^{1/2}\biggl\{I- U_tH_{int}^{-1/2}H_{s.a}H_{int}^{-1/2}U^*_t\biggr\}
H_{int}^{1/2}
\\
&\le
H_{int}^{1/2}\biggl\{I+\mu H_{int}^{-1/2}H_{int}^{\varepsilon }
H_{int}^{-1/2}\biggr\}
H_{int}^{1/2}
\\
&\le H_{int}+\mu(H_{int}+I)\le(1+\mu)\nu\Lambda +I\mu
\le c_0\Lambda.
\end{align*}
Thus under the above assumptions (2.3)
there exists a constant $c_0=\mu+(1+\mu)\nu$
such that $U_t\Lambda U^*_t\le c_0\Lambda$.

Assume that $\Lambda$ is a reference operator for some
formal generator ${\cal  L}(\cdot)$. Then we have
$\Phi(\Lambda)-G^*\Lambda-\Lambda G\le c\Lambda$.
Assume that
\begin{equation}
\Phi(H_{s.a.})\le c_1 \Lambda,\quad
\Phi(H_{int}^{\varepsilon })\le c_2 \Lambda,
\end{equation}
and consider an estimate the action of the formal generator
${\cal  L}(\cdot)$ in the interaction representation   o
n the element $\Lambda$:
\begin{align*}
{\cal  L}_t(\Lambda)&=
U_t^*\biggl(\Phi(U_t\Lambda U_t^*)-G^*\Lambda-\Lambda G\biggr)U_t
\\
&=U_t^*\biggl(\Phi(U_t(H_{int}-H_{s.a.}) U_t^*)
-G^*\Lambda-\Lambda G\biggr)U_t
\\
&=U_t^*\biggl(\Phi(H_{s.a.}-U_tH_{s.a.} U_t^*)+\Phi(\Lambda)
-G^*\Lambda-\Lambda G\biggr)U_t
\\
&\le U_t^*\biggl(\Phi(H_{s.a.}-U_tH_{s.a.} U_t^*)+c\Lambda\biggr)U_t
\\
&\le U_t^*\biggl(\Phi(H_{s.a})+\mu\Phi(H^{\varepsilon }_{int})+
c\Lambda\biggr)U_t
\\
&\le \bigl(c_1+\mu c_2+c\bigr)U_t^*\Lambda U_t
\le c_0(c_1+\mu c_2)+c)\Lambda.
\end{align*}
Thus, under the above assumptions
$$
{\cal  L}_t(\Lambda)\le \lambda \Lambda,\quad
\lambda=c_0\,(c+c_1+\mu c_2).
$$
By Theorem 2.1,
this estimate implies that the formal generator
$
{\cal  L}_t(X)=U_t^*{\cal  L}(U_t X U_t^*)U_t
$
is regular. On the other hand, the minimal quantum dynamical semigroup
generated by ${\cal  L}_t(\cdot)$ is unitary equivalent to the
minimal dynamical semigroup generated by
$$
\widetilde{\cal  L}(\cdot)={\cal  L}(\cdot)+i[H_{int},\cdot].
$$
Hence $\widetilde{\cal  L}(\cdot)$ also generates a unital
minimal dynamical semigroup, and it is regular too.
Its coefficients are $\Phi(\cdot)$ (the same CP-map),
and $H=H_s-\Lambda +H_{int}=H_s+H_{s.a.}$.
Thus one can add a self-adjoint operator
$H_{s.a.}$ to any regular generator ${\cal  L}(\cdot)$
which possesses a reference operator $\Lambda$ if
conditions (2.3)--(2.4) are fulfilled. In this case
we call $(\Lambda, H_{int})$ a $\Lambda$-{\sl pair} for the
generator $\widetilde{\cal  L}(\cdot)$.

\begin{thm}
Assume that the domain and continuity conditions
are fulfilled. If for a formal generator ${\cal  L}(\cdot)$
there exists a $\Lambda$-pair, the generator ${\cal  L}(\cdot)$
is regular.
\end{thm}
%

\section{Structure of generators of MME in quantum optics}
\setcounter{equation}{0}
The typical formal generator ${\cal  L}(\cdot)$ of a
Markov master equation in quantum optics
(see \cite{SBP96} (Schack \& Brun `96), \cite{TG96} (Brun \& Gisin `96),
\cite{AS97} (Ariano \& Sacchi `97), \cite{ZG97}
(Zoller \& Gardiner `97))
acts in ${\cal  B}({\cal  H})$,
${\cal  H}=(\ell_2)^{\otimes N}\otimes \C^M$;
its Lindbladian form  reads as follows:
${\cal  L}(B)=\Phi(B)-G^*B-BG,$
where
\begin{gather}
\Phi(B)=\sum_{k=1}^{N}\Phi_k(B),\quad
\quad  G=\frac{1}{2}\Phi(I)+iH,\notag
\\
\Phi_k(B)=\lambda_k a_k^\dagger B a_k \quad {\rm or}\quad
\Phi_k(B)=\mu_k a_k B a_k^\dagger,
\end{gather}
$\lambda_k, \, \mu_k\ge0$ are positive operators in ${\C^M}$,
$a_k$ and $a_k^\dagger$ are adjoint
creation and annihilation operators  acting on $k$-th factor of the
tensor product $(\ell_2)^{\otimes N}$, i.e.
($[a_k,a_n]=0,\quad [a_n,a^\dagger_k]=\delta_{k,n}$),
and
\begin{equation}
H=\sum_j \biggl(h_{j}
\prod_{k=1}^N (a_k^\dagger)^{n_{jk}}
(a_k)^{m_{jk}} +{\rm h.a.}\biggr)
\end{equation}
is an operator
in ${\cal  H}$ represented by a symmetric polynomial of a finite degree in
creation and annihilation operators with  matrix
coefficients $h_j\in \C^M\otimes \C^M$ (see
\cite{WV98} (Wiseman \& Vaccaro `98), \cite{KOBD98} (Kist, Orszag, Brun \&
Davidovich `99).

Note that the Hermitian structure (3.2) of the operator $H$ does
not imply its self-adjointness. For example, the Hamiltonian
of the third order
$$
H=-\frac{i}{\sqrt2}\biggl( (1+\frac{1}{2}(a+a^\dagger)^2)
(a-a^\dagger)+(a-a^\dagger)
(1+\frac{1}{2}(a+a^\dagger)^2) \biggr)
$$
is not a s.a. operator in $l_2$, because it is unitarily equivalent
to the symmetric operator
$\widehat H=i((1+x^2)\partial_x +\partial_x (1+x^2))$ in ${\cal L}_2(\R)$,
which has the nontrivial eigenvector
\begin{equation}
\psi(x)=\frac{\psi_0}{\sqrt{1+x^2}}e^{-\frac{1}{2}{\rm arctg(x)}}
\in {\cal L}_2(\R),\quad \psi_0\in \C
\end{equation}
such that $\widehat H\psi=-i\psi$. Hence the symmetric operators
$H$ and $\widehat H$ have  the same nontrivial deficiency index, and
$\pi=|\psi\rangle\langle\psi|$  is a projector to the deficiency
subspace $X={\cal  H}_d$.

Consider conditions on the projector $\pi$ which ensure
the violation of the unital property for equations with
constant operator coefficients.
We recall that the condition necessary and sufficient for
the minimal solution to be unital
is the weak convergence to $0$ of the monotone sequence of bounded
positive operators
\begin{equation}
Q_\varepsilon^n  (I)\to0,\quad
Q_\varepsilon  (X)
\stackrel{def}{=}\int_0^{\infty} \,dt\,e^{-\varepsilon t}V_{0,t}\Phi(X)
\end{equation}
(see \cite{Ch91}). Hence
the existence
of a positive bounded operator $X$, $\|X\|\le 1$, such that
$Q_\varepsilon (X)\ge X$ for some $\varepsilon >0$ is sufficient
for the violation of the unital property, since
the sequence
\begin{equation}
Q_\varepsilon^n (I)\ge Q_\varepsilon (X)\ge X\ge 0
\end{equation}
clearly does not converge to $0$.

%
\begin{thm}
If the coefficients of a formal generator ${\cal  L}(\cdot)$
satisfy domain and continuity assumptions  and there exist a positive
bounded operator $X$, $||X||\le1$, and $\varepsilon>0 $ such that
\begin{equation}
{\cal  L}_*(X)[\psi]\ge \varepsilon X_*[\psi]\quad \forall
\psi\in\dom G^N={\cal  D}_N,
\end{equation}
then the corresponding minimal quantum dynamical semigroup
does not preserve the unit operator.
\end{thm}

\proof
Let us derive the inequality
$Q_\varepsilon(X)\ge X $ from (3.6).
 Inequality (3.6) implies that
      $\Phi(X)_*\ge
      (\varepsilon X+G^*X+XG)_*$ on ${\cal  D}_N$.
      Then for  $\psi \in {\cal  D}_N$, we have
      $\psi_t = W_t\psi \in {\cal  D}_N$ and
\begin{align*}
e^{-\varepsilon t}\Phi_*(X)[\psi_t]&\ge
e^{-\varepsilon t}\bigl(\varepsilon
(\psi_t,X\psi_t) +(G\psi_t,X\psi_t)
+(X\psi_t,G\psi_t)\bigr)\\
&=-\frac{d}{dt}
\bigl(e^{-\varepsilon t}\|X\psi_t\|^2\bigr).
\end{align*}
      Therefore,
$$
\int_0^t e^{-\varepsilon \tau}\Phi(X)_*[\psi_{\tau}]\,d\tau
\ge
(\psi,X\psi)-e^{-\varepsilon t}(\psi_t,X\psi_t).
$$
     The limit as $t\to \infty$ yields an inequality for $X$:
     $(\psi,Q_\varepsilon(X)\psi)\ge (\psi,X\psi)$
     for any $\psi\in {\cal  D}_N$
     by definition of the map $Q_\varepsilon (\cdot)$.
Since ${\cal  D}_N$ is dense in $\ch$ and the map $Q_\varepsilon(\cdot)$
is bounded, this inequality is equivalent to
\begin{equation}
Q_\varepsilon(X)\ge X,
\end{equation}
which contradicts the necessary unitality condition.
\endproof

A natural candidate to be used as $X$ in inequality (3.6) is
 the  projector to the deficiency subspace
of the operator $H$, if such a subspace exists.

\begin{thm}
{\rm  \cite{CS00} (Chebotarev \& Shustikov `00)}

Let $H$ be a densely defined symmetric operator.
Assume that it has a nontrivial deficiency
subspace $\ch_d=\{\psi:\, H^*\psi=-i\psi\}$
and $\pi_d$ is the projection onto $\ch_d$.
If moreover,
there exists $\varepsilon>0 $ such that
\begin{equation}
\Phi(\pi_d)_*[\psi]-\re (\Phi(I)\psi,\pi_d\psi)\ge
-(2-\varepsilon )\|\pi_d\psi\|^2
\end{equation}
for all $\psi\in {\cal  D}_N$,
then inequality {\rm (3.6)} holds and the necessary
unitality condition
{\rm (3.4)}
is violated.
\end{thm}

\proof
Let us prove that (3.8) implies (3.6) in the following form:
$$
{\cal L}(\pi_d)_*[\psi]\ge \varepsilon \|\pi_d\psi\|^2\quad
\forall \psi\in {\cal  D}_N.
$$
Indeed,  since
$H^*\pi_d=-i\pi_d$ and $\pi_d=\pi_d^2$,
it follows from (3.8) that
\begin{gather}
{\cal L}(\pi_d)_*[\psi]={\Phi}(\pi_d)_*[\psi]-\re(\Phi(I)\psi,\pi_d\psi)+
i\bigl( (H\psi,\pi_d\psi)-(\pi_d\psi,H\psi) \bigr)
\notag\\
={\Phi}(\pi_d)_*[\psi]-\re(\Phi(I)\psi,\pi_d\psi)+2||\pi_d\psi||^2\ge
\varepsilon \|\pi_d\psi\|^2
\end{gather}
and inequality (3.6) is true.
\endproof

Note that for any finite polynomial $H=H_2+H_{s.a.}$ in creation and
annihilation operators $a_k^\dagger$ and $a_k$,
there exists a diagonal operator
\begin{equation}
\Lambda=c_{\Lambda}\biggl(1+\sum_{k=1}^N (a_k^\dagger a_k)^{m_k}\biggr),
\quad c_{\Lambda}>0
\end{equation}
such that $H_{2}$ and $H_{s.a}$ are
relatively bounded by $\Lambda$ with  the relative upper bound
$O(c_\Lambda^{-1})$.
One can use $\Lambda$ as  the  reference operator.
In any case we assume that $\Phi(I)\ge I$ is a s.a. operator
and
$$
a\,\dom\Lambda\subseteq\dom\Lambda^{1/2},\quad
a^\dagger\,\dom\Lambda\subseteq\dom\Lambda^{1/2}.
$$
The last two assumptions readily hold if $m_k\ge M=2$ in (3.10).

\begin{thm}
If the Hamiltonian $H$ can be represented
as $H=H_2+H_{s.a.}$, where $H_{s.a.}$ is a self-adjoint polynomial
of a finite order $M$ in creation and annihilation  operators
and $H_2=H_2(a^\dagger,a)$ is a polynomial  of the second order,
then there exist $c_\Lambda>0$ and $\{m_k\}\ge M$
such that
$
(\Lambda,\, H_{int}=H_{s.a.}+\Lambda)
$
is a $\Lambda$-pair for the generator {\rm (3.1)--(3.2)}.
\end{thm}

\proof
We recall that any finite polynomial in creation
and annihilation
operators of order $M$ can be dominated by the diagonal operator
(3.10) of higher order,
provided the constant $c_D$ is sufficiently
large and $ N=\min\{m_k\}\ge M$. Hence for sufficiently large $c_D$,
and $N$
by the classical perturbation theory  \cite{K76},
$H_{int}=H_{s.a.}+\Lambda$ is a positive s.a. operator such that
$\dom H_{int} =\dom\Lambda$, and
$G_0=i(H_2-\Lambda)+\Phi(I)/2$ is an accreative operator,
$\dom G_0=\dom\Lambda$. Since $\Phi(\cdot)$ and $H_{s.a.}$ are operators
of a finite (second) order, the property
(2.3) and (2.4) of $\Lambda$-pair can readily be fulfilled by
choosing $M,\,N$ and $c_\Lambda$ sufficiently large.

The commutator of a polynomial of the second order in creation
and annihilation operators with arbitrary polynomial
of order $M<\infty$ has the order $M$ or less. Hence, the commutator
$$
i[H_0,\Lambda]=i[H_2-\Lambda,\Lambda]=i[H_2,\Lambda]
$$
is an operator of the same order as $\Lambda$, and hence there exists a
constant $c\in \R$
such that $i[H_0,\Lambda]\le c\Lambda$.

A simple algebra shows that for CP-map (3.1), the operator
$\Phi(\Lambda)-(\Lambda\Phi(I)+\Phi(I)\Lambda)/2$
is also a polynomial of the same order as $\Lambda$.
More precise,
the following two estimates hold:
$$
(a_k^\dagger)^l  \Lambda a^l_k-\frac{1}{2}\biggl(
(a_k^\dagger)^l a_k^l  \Lambda
+ \Lambda (a_k^\dagger)^l a_k^l\biggr)\le 0,\quad
$$
for any $l\ge0$, and on the other hand there exists $c\ge 0$ such that
$$
a_k  \Lambda a_k^\dagger-\bigl(a_k a_k^\dagger  \Lambda
+ \Lambda a_ka_k^\dagger \bigr)/2\le c\Lambda,
$$
for the operator $\Lambda$ (3.10).
Therefore, there exists $c>0$ such that
${\cal  L}_0(\Lambda)\le c \Lambda$ on $\dom\Lambda$,
and hence  $\Lambda$ is a reference operator
for
$$
\Phi_{k,l}(B)=\lambda_{k,l}(a_k^\dagger)^lBa_k^l,\quad
\Phi_{k}(B)=\lambda_{k}a_kBa_k^\dagger.
$$
This proves the theorem.
\endproof

\setcounter{equation}{0}
\section{Examples}
In this section we consider some classes of Hamiltonians
and completely positive maps for which our Theorem 2.2
is applicable.

1. Let $\lambda$ be a complex number and $m,n\ge0$. Set
\begin{equation}
H=\lambda(a_1^\dagger)^{m} a_2^{n}+
\overline \lambda a_1^{m}(a_2^\dagger)^{n},\quad \lambda\in\C.
\end{equation}
Let us prove that all Hamiltonians of such form are essentially self-adjoint
in ${\cal H}_2=l_2\otimes l_2$. It suffices to
prove that there does not exist
a vector
$$
\psi=\{\psi_{k,j},\;k,j\ge1,\;
\sum_{k,j}|\psi_{k,j}|^2=||\psi||^2_{{\cal H}_2}\}
$$
such that $H\psi=\pm i\psi$ \cite{RS}. We set $\psi_{k,j}=0$
if $\min\{k,j\}\le0$.

Let us rewrite these equations  for components $\psi_{k,j}$ as follows:
\begin{equation}
\pm i \psi_{k,j}=\lambda A_{k,j}^{m,n}\psi_{k-m,j+n}+
\overline \lambda B_{k,j}^{m,n}\psi_{k+m,j-n},\quad
A_{k,j}^{m,n},\;B_{k,j}^{m,n}\ge0.
\end{equation}
We set $\psi_{k,j}=0$
if $\min\{k,j\}\le0$ and
skip exact expressions for the functions $A_{k,j}^{m,n}$ and
$B_{k,j}^{m,n}$ because they are
irrelevant for the proof.
The important property of this system is
that it splits into a set of {\sl independent finite }
subsystems of linear algebraic
equations with respect to values of one of the components of the set
$$
X_k=\{x_j=\psi_{k-jm,1+jn},\;k-jm\ge1,
\; jn\ge 1,\; j=0,1,\dots,[k/m]-1 \},
$$
where $x_j=0$ for all $j<0$.
For each $k,m,n$ fixed, the system of linear algebraic equations
corresponding to (4.2) has
 the  three-diagonal form
$$
\pm ix_j=\lambda A_j x_{j+1}+ \overline \lambda B_j x_{j-1}
$$
with some positive $A_j$ and $B_j$. But it is a well-known fact
(see \cite{Be}) that
$$
D_N=\det\left(\begin{array}{cccccc}\pm iI&\lambda A_1&0&\dots&0&0\\ \\
\lambda^* B_1&\pm iI&\lambda A_2&\dots&0&0\\ \\
{}&{}&{}&{\dots}&{}&{} \\ \\
0&0&0&\dots& \lambda^* B_{N-1}&\pm iI
\end{array}\right)\ne 0
\eqno(4.3)
$$
where the entries of the matrix are $(k\, \times \,k)$-blocks, $\lambda$ and
$\lambda^*$ are Hermitian adjoint $(k\, \times \,k)$--matrices,
and $I$ is the unit
matrix in $\C^M$.
\setcounter{equation}{3}

By the Gershgorin theorem \cite{Be}, Hamiltonians (4.1) are
relatively boun\-ded
by the diagonal matrix $\Lambda$ (2.3) of order $M\ge m+n$, and the
relative upper bound decreases as $c_D\to\infty$. Hence all
formal generators with the completely positive parts (3.1) and Hamiltonian
part
(4.1) are regular.

\medskip

2. The same assertion is true for Hamiltonians
from ${\cal C}(l_2^{\otimes N}\otimes \C^M)$ of the following form:
$H=H_{int}+H_0$,
\begin{equation}
H_{int}=
\lambda(a_1^\dagger)^{m_1}a_1^{n_1}\cdots a_N^{m_N}(a_N^\dagger)^{n_N}+
\lambda^* a_1^{m_1}(a_1^\dagger)^{n_1}\cdots(a_N^\dagger)^{m_N}a_N^{n_N},
\end{equation}
where $\lambda$ and $\lambda^*$ are Hermitian adjoint
$(M\times M)$-matrices, $ \sum m_k+n_k=K,$ and $H_0$
is any symmetric operator dominated by $\Lambda$
and such that
$$
\exists c\in \R:\quad i[H_0,\Lambda]\le c\Lambda.
$$
The proof of self-adjointness of $H_{int}$ is based on a similar
factorization of the set of block-matrices
$\{\psi_{k_1,\dots k_N}\}\in l_2^{\otimes N}\otimes \C^M $ and
on the reduction of the homogeneous system of linear algebraic equations to
the set of finite-dimensional linear equations with
nondegenerate three-diagonal $(M\times M)$-block matrix (4.3).

As in the previous case, the interaction
representation is generated by the self-adjoint operator $H_{int}$
dominated by the diagonal operator $D$ for $M\ge \sum(m_k+n_k)$,
and the Hamiltonian
$H_0$ of ${\cal L}_0(\cdot)$, because it satisfies
the conservativity and compatibility conditions. For generators (1.1),
any symmetric operator  on $H_0$ of the second order in creation and annihilation
satisfies the above assumptions.

\medskip

3. Consider the physical example \cite{SBP96} (Schack, Brun \& Pecival `96)
of a formal generator ${\cal L}(\cdot)$
in ${\cal B}(l_2\otimes\l_2\otimes\C^2)$
 with CP-part (1.1) and the Hamiltonian
\begin{equation}
\hat H = Ei( a_1^\dagger- a_1) + \frac{\chi}{2}i( a_1^{\dagger2} a_2
  - a_1^2 a_2^\dagger) + \omega\sigma_+\sigma_-
  + \eta i( a_2\sigma_+ - a_2^\dagger\sigma_-),
\end{equation}
where $E$ is the strength of an external pump field, $\chi$ is the strength of
the interaction, $\omega$ is the detuning between the frequency of
the field mode $ a_2$ and the spin transition frequency, and $\eta$ is
the strength of the coupling of the spin to the field mode $ a_2$. The
completely positive part of the generator reads as follows
\begin{equation}
\Phi(B)={2\gamma_1} a_1^\dagger B a_1+{2\gamma_2}a_2^\dagger B a_2
+{2\kappa}\,\sigma_+ B\sigma_-.
\end{equation}
It describes the dissipation of the field modes
and the spin with coefficients
$\gamma_1$, $\gamma_2$, and $\kappa$, respectively; $\sigma_\pm$ are
two by two matrices.

The Hamiltonian (4.5) can be readily represented in the form
(4.3) with $k=2$, $K=3$, $\lambda=I$,
$H_{int}=\frac{\chi}{2}i( a_1^{\dagger2} a_2
  - a_1^2 a_2^\dagger)$ and $H_0=H-H_{int}$. The completely positive
part has the form (3.1). Hence the formal generator (4.5)--(4.6) is
regular.

\medskip

4. The kinetic stage of the evolution of a quantum system
interacting with environment is described in \cite{KS97}
(Kilin \& Schreiber `97)
by the following Markov master equation:
\begin{align*}
\frac{\partial \sigma}{\partial t}
&=-i\omega \left[ H(a^{\dagger},a),\sigma \right] \label{12a}
\\
  &+ \Gamma_2 (n_2 + 1)
     \left\{
       \left[
          a^2
           \sigma,
        (a^{\dagger})^2
       \right]
      + \left[
        (a^{\dagger})^2,
        \sigma
          a^2
       \right]
     \right\}
\\
 &+ \Gamma_2 n_2
     \left\{
       \left[
          (a^{\dagger})^2 \sigma, a^2
       \right]
      + \left[
        a^2, \sigma (a^{\dagger})^2
       \right]
     \right\},
\end{align*}
where s.a. operator $H= H(a^{\dagger},a)$
is a finite symmetric polynomial in $a^{\dagger}$ and $a$ of order
no greater 4,
$ \Gamma_2 =\pi K^2 g_2 $ is the decay rate of the vibrational
amplitude.
Here, the number of quanta in the bath mode
$n_2=n(2 \omega)$,
the coupling function
$K=K(2 \omega)$,
and the density of bath states
$g_{2}=g(2\omega)$ are evaluated at the double frequency
of the selected oscillator.
The corresponding dual CP-map $\Phi(\cdot)$ acts as follows
$$
\Phi(B)=2\Gamma_2\biggl( (n_2+2)(a^\dagger)^2 Ba^2+n_2 a^2B(a^\dagger)^2
\biggr).
$$
This case is rather simple: $\Lambda=c (a^\dagger)^2a^2$ and
$H_{int}=H+\Lambda$, where $c$ is sufficiently large:
$\lambda\ge \Phi(I)$, and $||\Lambda h||\ge 2||Hh||$, so that
$H_{int}$ is a s.a. operator, provided $H$ is s.a. operator.
Hence the generator of the above master equation is regular.

\medskip

5. The previous example can be generalized as follows. Set
$$
\Phi^+_m(B)=(a^\dagger)^mBa^m,\quad  \Phi^-_m(B)=a^mB(a^\dagger)^m,
\quad \Lambda_n=((a^\dagger a)^n+I)\lambda,
$$
$\lambda>0$. Then we set
$$
\Lambda_n\psi_N=\lambda(N^n+1)\psi_N,\quad
\Phi^\pm_m(I)\psi_N= \left(\frac{(N\mp m)!}{N!}\right)^{\mp1}\psi_N
$$
for $N$-particle component of $\psi_N$ of the vector
$\psi=\{\psi_0,\psi_1,\dots\}\in l_2$.
Hence there exists $n\ge m$ and $\lambda=\lambda(m,n)>0$ such that
$\Lambda_n\ge\Phi^\pm_m(I)$.

Similarly we obtain
\begin{gather*}
\biggl(\Phi^\pm_m(\Lambda_n)-\bigl(\Lambda_n\,\Phi^\pm_m(I)
+\Phi^\pm_m(I)\Lambda_n\bigr)/2\biggr)\psi_N=
\\
= \lambda\left(\frac{(N\mp m)!}{N!}\right)^{\mp1}\bigl((N\mp m )^n-N^n\bigr)
\psi_N.
\end{gather*}
Therefore, for any formal generator with completely positive part
$$
\Phi(B)=\sum_k\bigl\{c_k^+\Phi_{m_k}^+(B)+c_k^-\Phi_{n_k}^-(B)\bigr\}
$$
with positive matrix coefficients $c_k^\pm\in \C^M\otimes\C^M$,
the third $\Lambda$-pair assumption is fulfilled if
the {\it balance condition} is true:
\begin{gather*}
\sup_{N\ge1}\sum_k\biggl\{
c_k^+\frac{N!}{(N-m_k)!}\bigl[(1-m_k/N)^n-1\bigr]+
\\
+c_k^-\frac{(N+n_k)!}{N!}\bigl[(1+n_k/N)^n-1\bigr]\biggr\}<cI .
\end{gather*}
Then the regularity conditions of Theorem 2.2 are fulfilled for
$\lambda$ sufficiently large if $H=H_2+H_{s.a.}$, where
$H_2(a,a^\dagger)$ is any
Hermitian quadratic polynomial, and
$H_{s.a.}(a,a^\dagger)$ is any self-adjoint Hamiltonian
of order less or equal $2n$.
The balance condition is readily fulfilled if
$$
S=\sum_k\bigl\{ c_k^+m_k-c_k^-n_k\bigr\}>0
$$
is a strictly positive operator in $\C^M\otimes\C^M$.

The generators of MME with completely positive component of the fourth
order was used in \cite{SM97} (Schneider \& Milburn `97):
\begin{equation}
{\cal  H}=l_2\otimes \C^2,\quad
\Phi(B)=(a^2\sigma_++(a^\dagger)^2\sigma_-)B(a^2\sigma_-
+(a^\dagger)^2\sigma_+),
\end{equation}
where
$$
\sigma_+=\left(\begin{array}{cccc}0&1\\0&0\end{array}\right),\quad
\sigma_-=\left(\begin{array}{cccc}0&0\\1&0\end{array}\right).
$$
 Hence, for diagonal   operators $B\in{\cal  B}({\cal  H})$
$$
B=\left(\begin{array}{cccc}B_1&0\\0&B_2\end{array}\right),
\quad B_{1,2}\in {\cal  B}(l_2)
$$
we have $\sigma_\pm^2=0$ and
$$
\Phi(B)=\left(\begin{array}{cccc}a^2B_1(a^\dagger)^2&0
\\0&(a^\dagger)^2B_2a^2\end{array}\right).
$$
Therefore, the generator ${\cal  L}(\cdot)$ has  the  component
$$
{\cal  L}_{11}(X)=a^2X(a^\dagger)^2-(a^2(a^\dagger)^2X+X(a^\dagger)^2)/2+
i[H_{11},X]
$$
which is unregular for any first order operator $H_{11}$.

\medskip

6. The paper \cite{LMV97}  (Lanz, Melsheimer \& Vaccini `97)
presents examples of formal generators in
${\cal  H}=l_2\otimes l_2$ with the CP-component
$
\Phi(B)=a_1\,a^\dagger_2\,B\,a_2\,a_1^\dagger.
$
Let us prove that for the generators with coefficients
\begin{equation}
\Phi(B)=a_1^L(a^\dagger_2)^MBa_2^M(a_1^\dagger)^L,
\quad L,M\ge 1,\quad H=H_2+H_{s.a.}(a^\dagger,a)
\end{equation}

there exists a $\Lambda$-pair. Consider the generator
$$
{\cal  L}_0(B)=a_1^L(a^\dagger_2)^MBa_2^M(a_1^\dagger)^L-
B\circ a_1^L(a_1^\dagger)^L(a^\dagger_2)^Ma_2^M.
$$
Straightforward computation proves that $\Phi(I)$ is not a reference
operator for ${\cal  L}_0(\cdot).$

\begin{lem}
For any $N\ge0$, there exists a polynomial
\begin{equation}
\Lambda_N=\lambda^{(N)}I+\sum_{k=0}^N\lambda_k^{(N)}
(a_1^\dagger)^{N-k}a_1^{N-k}(a_2^\dagger)^{k}a_2^k,\quad
\lambda^{(N)},\,\lambda_{k}^{(N)}\ge1
\end{equation}
and a real constant $c_N$
such that ${\cal  L}_0(\Lambda_N)\le c_N\Lambda_N$.
\end{lem}

\proof
Note that
\begin{align*}
{\cal  L}_0((a_1^\dagger)^{m}a_1^{m}(a_2^\dagger)^{n}a_2^n)&=
mL(a_1^\dagger)^{L+m-1}a_1^{L+m-1}(a_2^\dagger)^{n+M}a_2^{n+M}
\\
&-nM(a_1^\dagger)^{L+m}a_1^{L+m}(a_2^\dagger)^{n+M-1}a_2^{n+M-1}+{\rm
l.o.t.},
\end{align*}
where the lower order terms (l.o.t.) can be dominated by the main terms and
$\lambda^{(N)}I$ for all $\lambda^{(N)}$ sufficiently large.
Hence,
\begin{align*}
{\cal  L}_0(\Lambda_N)&=\sum_{k=0}^{N-1}[L(N-k)\lambda^{(N)}_k-
M(k+1)\lambda^{(N)}_{k+1}]
\\
&\times (a_1^\dagger)^{L+N-k-1}a_1^{L+N-k-1}(a_2^\dagger)^{k+M}a_2^{k+M}+
{\rm l.o.t.}
\end{align*}
Therefore, all main terms have {\it negative} coefficients if
\begin{equation}
\lambda_{k+1}^{(N)}>\lambda_{k}^{(N)}\frac{(N-k)L}{(k+1)M}.
\end{equation}
For the fixed $N$, the lower order terms can be dominated by
the main terms plus
$\lambda^{(N)}I$ for all $\lambda^{(N)}$ sufficiently large,
that is
$$
\exists c>0:\quad
{\cal  L}_0(\Lambda_N)\le c\lambda^{(N)}I\le c\Lambda_N
$$
if (4.10) holds. Since
$$
\Lambda_N\ge\lambda^{(N)}I+\lambda_0^{(N)}(a_1^\dagger)^Na_1^N+
\lambda_N^{(N)}(a_2^\dagger)^Na_2^N,
$$
where the coefficients $\lambda^{(N)},\,\lambda_0^{(N)},\,
\lambda_N^{(N)}$ can be chosen greater than any constant $c\ge 0$.
In particular, they can be chosen such that
the  diagonal  s.a. operator $\Lambda_N$ dominates
 with arbitrary small upper relative bound
a given polynomial $H_{s.a.}$. Note that for any quadratic
operator $H_2=H_2(a^\dagger,a)$, ${\cal  L}_0(H_2)$ is a symmetric
polynomial
of order $2(L+M)$ in creation and annihilation operators. Hence it can
be dominated by $\Lambda_N$ for any $N\ge 2(L+M)$.
In this case,
$
\{\Lambda_N,\;H_{int}=\Lambda_N+H_{s.a.}\}
$
is a $\Lambda$-pair for the formal generator ${\cal  L}(\cdot)$
with coefficients
(4.8). This proves that ${\cal  L}(\cdot)$ is regular.
\endproof

\section{Discussion}
By using the concept of $\Lambda$-pair, we have analyzed
the regularity property for a wide class of  generators of MME in quantum
optics which are available for authors. We proved that the
generators of the form
$$
{\cal  L}(B)=\Phi(B)-(\Phi(I)B+B\Phi(I))/2+i[H_2+H_{s.a.},B]
$$
are regular for CP-maps (3.1), (4.6), and (4.8) if
$H_{s.a.}$ is a self-adjoint polynomial of a finite order
in creation and annihilation operators, and $H_2$ is a symmetric
operator of the second order.
To conclude the paper, we recall the most important open problems.

Generator of MME  can be irregular    if $\Phi(\cdot)$ is as in (4.7).
From mathematical viewpoint, to select a unique solution,
one must introduce a kind of boundary
condition as was done in \cite{Ch94}, where all unital extensions
of the minimal quantum dynamical semigroup are described in terms
of extension of its resolvent. In analogous classical cases,
the boundary conditions for stochastic processes follow from Dynkin's
formula \cite{Dy65}, \cite{Gr99} for infinitesimal operator
of the Markov semigroup.
The physical sense of boundary conditions
for quantum systems should be related to conservation laws,
but physical  examples of MME with boundary conditions still
are not known.

\vskip 2pc
{\obeylines
\noindent Quantum Statistics Dept.,
\noindent Moscow State University,
\noindent Moscow 119899, Russia
\noindent e-mail: {\tt alex\@@cheb.phys.msu.su}
}
\vskip 1pc
{\obeylines
\noindent Dipartimento di Matematica,
\noindent Univ. Autonoma Metropolitana,
\noindent Iztapalapa,  Mexico D.F.,  09340,
\noindent e-mail: {\tt jcgc\@@xanum.uam.mx}
}

\vskip 1pc
{\obeylines
\noindent Dipartimento di Matematica,
\noindent Univ. Autonoma Metropolitana,
\noindent Iztapalapa,  Mexico D.F.,  09340,
\noindent e-mail: {\tt  roqb\@@xanum.uam.mx}
}
\end{document}